\documentclass[useAMS,usenatbib,dcolumn,usegraphicx,twocolumn,10pt]{mnarxiv}
\usepackage{times}
\usepackage[active]{srcltx}
\usepackage{placeins}
\newcommand{\ud}[1]{\mathrm{d}#1}

\newcommand{\vp}{\mathrm{\textit{V p}}}
\newcommand{\br}[1]{\left(#1\right)}
\newcommand{\abs}[1]{\left|#1\right|}
\renewcommand{\sq}[1]{\left[#1\right]}
\newcommand{\eqref}[1]{(\ref{#1})}

\newcommand{\kpc}{\mathrm{kpc}}
\newcommand{\pc}{\mathrm{pc}}

\newcommand{\msun}{\mathrm{M}_{\sun}}
\newcommand{\gu}{\mathrm{\ km\ s^{-1}\ kpc^{-1}}}
\newcommand{\sgn}[1]{\mathrm{sgn}\br{#1}}
\title[]{Transverse gradients of azimuthal velocity in a global disc model of the Milky Way Galaxy}

\author[]{Joanna Ja{\l}ocha$^{1}$,
{{\L}ukasz Bratek$^{1}$},
{Marek Kutschera$^{1,2}$,}
{Piotr Skindzier$^{2}$}
\\
$^{1}$Institute of Nuclear Physics,
Polish Academy of Sciences, Radzikowskego 152, PL-31342 Krak\'{o}w, Poland\\
$^{2}$Institute of
Physics, Jagellonian University,  Reymonta 4, PL-30059 Krak{\'o}w, Poland}

\begin{document}
\date{\today}
\pagerange{\pageref{firstpage}--\pageref{lastpage}} \pubyear{2009}

\maketitle

\begin{abstract}
  In this paper, we aim to estimate the vertical gradients in the rotational velocity of the Galaxy. This is carried out in the framework of a global thin disc model approximation. The predicted gradient values coincide with the observed vertical fall-off in the rotation curve of the Galaxy. The gradient is estimated based on a statistical analysis of trajectories of test bodies in the gravitational field of the disc and in an analytical way using a quasi-circular orbit approximation. The agreement of the results with the gradient measurements is remarkable in view of other more complicated, non-gravitational mechanisms used for explaining the observed gradient values. Finally, we find that models with a significant spheroidal component give worse vertical gradient estimates than the simple disc model. In view of these results, we can surmise that, apart from the central spherical bulge and Galactic halo, the gross mass distribution in the Galaxy forms a flattened rather than spheroidal figure.\\

\medskip
\hrule
\flushleft \textbf{The definitive version is available at \\ \texttt{http://onlinelibrary.wiley.com/\\doi/10.1111/j.1365-2966.2010.16987.x/abstract}}
\medskip
\hrule
\end{abstract}


\section{Introduction}
Recently, apart from other local and global characteristics of the rotation speed of the Galaxy, such as rolling motion, \citet{bib:gradients} determined the fall-off rate $\gamma=-22\pm6\gu$ in the rotation speed from the Galactic mid-plane. This gradient estimation was obtained by fitting a linear profile\footnote{Throughout this paper we use cylindrical
coordinates $(r,\varphi,z)$ distinguished by the galactic mid-plane $z=0$ and the axis of rotation $r=0$.}
\begin{equation}\label{eq:model_rot}v_{\varphi}(r,z)=v_{\varphi}(r,0)+\gamma \abs{z}+\delta z\end{equation}
to rotation measurements in the vicinity of the Galactic mid-plane: $\abs{z}<0.1\,\kpc$,   $r\in\br{3,8}\,\kpc$.

 The principal purpose of our work is to reconstruct the vertical gradient magnitude in a simple model of the Galaxy and to find out to what extent the gradient behaviour is dependent on the geometry of mass distribution. In particular, the gradient value can be very well reconstructed in the global thin disc model, provided the disc comprises gross dynamical mass ascertained from the rotation curve of the Galaxy.

\subsection{Motivation of the present work}
There is a suggestive heuristics behind the above simple ansatz for the rotation speed. This naturally leads us directly to the global disc model approximation as a mean of determining the vertical gradient of the rotation speed in flattened galaxies, in particular, in the Galaxy.

It is a simple matter to note that every axisymmetric function $f(r,z)$ in the cylindrical coordinate frame can be written (for $z\ne0$) as
  \begin{equation}\label{eq:f}f(r,z)=f(r,0)+
\Gamma(r,z) \abs{z}+
\Delta(r,z) {}z\end{equation} Here, $\Gamma$ and $\Delta$ are $z$-symmetric functions defined as
\[\begin{array}{@{}l@{\,\,\,}r@{}}\Gamma(r,z)=\frac{f(r,\abs{z})+f(r,-\abs{z})-2f(r,0)}{2\abs{z}},&
\Delta(r,z)=\frac{f(r,\abs{z})-f(r,-\abs{z})}{2\abs{z}}\end{array}.\]
Consider now the azimuthal velocity $v_{\varphi}$ in place of the function $f$, $f(r,z)=v_{\varphi}(r,z)$.
 We expect $v_{\varphi}(r,z)$ to be nearly $z$-symmetric. In this case, the difference  $f(r,\abs{z})-f(r,-\abs{z})$ is small compared to the symmetric part $f(r,\abs{z})+f(r,-\abs{z})\approx 2f(r,z)$. From the Lagrange mean value theorem, it then follows that $\Gamma\approx\frac{f(r,z)-f(r,0)}{\abs{z}}=f_{,\zeta}(r,\zeta)|_{\zeta=hz} \sgn{z}$, $0<h<1$ and ${\Delta}\approx0$. When, in addition, $v_{\varphi}(r,z)$ falls off almost linearly with distance off the mid-plane and does not change significantly with $r$ in the considered region, then $\Gamma$ is nearly constant. In this case, we can write $v_{\varphi}(r,z)\approx{}v_{\varphi}(r,0)+\gamma\abs{z}$, where $\gamma$ is a constant such that $\gamma\approx\Gamma$. Thus, to the first order in $z$, constant $\gamma$ measures the fall-off rate or the vertical gradient of the speed of rotation. In the same order of approximation, we can include the correction from the antisymmetric part. Then $v_{\varphi}(r,z)\approx{}v_{\varphi}(r,0)+\gamma\abs{z}+\delta z$, provided $\Delta$ is also almost constant in the considered region. Under these conditions, we can use model \eqref{eq:model_rot} to estimate the characteristic magnitudes of $\Gamma$ and $\Delta$. Constant $\delta$ is then called a rolling parameter, as $\abs{\delta{z}}$ measures the difference of speeds on opposite sides of the galactic mid-plane. In this approximation, parameters $\gamma$ and $\delta$ are global characteristics of the rotation speed in the considered region of coordinates $r$ and $z$. Of course, the linear approximation can also be used when functions $\Gamma$ and $\Delta$ are not almost constant in this region. Then parameters $\gamma$ and $\delta$ characterize, on average, the fall-off rate and rolling in this region.

 In the $z$-symmetric case, customarily assumed in simple modelling of mass distribution in galaxies, we necessarily have $\Delta\equiv0$. Thus, the rolling parameter cannot be determined by such models, but the fall-off rate can be still determined. If we can show in such a galaxy model that $\Gamma$ indeed is almost constant in the considered region and its predicted value is correct, then we can regard the model of mass distribution as consistent to the first order of approximation with the method used to determine $\gamma$ from observations of rotation in the Galaxy based on the linear fit \eqref{eq:model_rot}. The same applies when $\Gamma$ is not almost constant; then, $\gamma$ characterizes, on average, the fall-off rate in the considered region. What counts is that the linear fit to observations \eqref{eq:model_rot} and a model value of the fall-off in a linear approximation are consistent with each other.

 In this paper, we attempt to realize the above programme of finding a simple model, which in the linear approximation would be able to predict the correct value of $\gamma$. Although expression  for $f(r,z)$ on the right-hand side of equation \eqref{eq:f}, which we used above as a starting point to show that approximation \eqref{eq:model_rot} is well grounded, is always valid -- has the same analytical properties as $f(r,z)$-- the approximate expression \eqref{eq:model_rot} is not analytical at $z=0$ as $\abs{z}$ cannot be differentiated. This is the same type of singularity that is encountered in the thin disc model in which the gravitational field is everywhere smooth apart from the plane $z=0$; the same goes for rotation on almost circular orbits close to this plane. The thin disc model is thus naturally compatible with approximation \eqref{eq:model_rot}. Of course, mass distribution in a real flattened galaxy does not form an infinitely thin disc and $v_{\varphi}\br{r,z}$ is a smooth function of which equation \eqref{eq:model_rot} is a cusp-like approximation. However, a more realistic model of mass distribution (say, finite width disc) in the same linear approximation should lead to similar results, although in an unnecessarily more complicated way. We have therefore decided to model vertical gradients in the rotational velocity of the Galaxy in a global thin disc model approximation. Later, we compare predictions of the model with the observed gradient value. We also find from numerical simulations of test bodies in the gravitational field of the disc that the vertical gradient is almost constant. This again gives further support that our model is consistent with the gradient measurement method \eqref{eq:model_rot} (constancy of the fall-off rate is also in accord with observations of the gradient in other galaxies).

 When the fall-off rate in the considered region is observed to be constant (which is the case for other galaxies), the important feature of linear approximation is that, in order to model the fall-off rate $\gamma$ in this region, we can investigate the behaviour of the model speed not necessarily very close to the galactic mid-plane, but also at heights greater than the stellar disc width. In this situation, the scaleheight should not be important as additional non-linear terms, which could be added to the fitting profile \eqref{eq:model_rot}, would be less significant. In other words, when the gradient value is observed to be constant, we should expect that models with intrinsic scaleheight (such as finite width discs) or without such scale (such as a thin disc) would predict comparable estimations for the gradient value in the region.

\citet{bib:gradients}
found the observed gradient values to be roughly four times greater than those predicted by a finite width exponential disc of stellar mass. They correctly concluded that to explain the gradient value some other process besides simple gravitational physics must have been taken into account. It is important to note that this situation cannot be remedied by taking into account other mass components, such as the dark halo, because close to the Galactic mid-plane any spheroidal mass component gives negligible contribution to the total gradient value. It vanishes on the mid-plane and only becomes significant further away (this issue is illustrated in more detail in the next paragraph). This is the principal difference between contributions of a spheroidal and a flattened mass distribution to the overall gradient value close to the Galactic mid-plane.

Interestingly, we show that the high gradient values could be explained by simple gravitational physics if only the premise about mass distribution in Galaxy were changed. To this end, we have assumed that perhaps except for the central core, the whole dynamical mass forms a flattened, disc-like, rather than spheroidal, structure. This hypothesis enables us to use a global thin disc model for the approximate description of a flattened mass distribution comprising all dynamical mass (i.e. that inferred from Galaxy rotation). We can therefore say that high gradient values may indicate simply that gross dynamical mass forms a flattened mass distribution, and not that other processes besides simple gravitational physics would be needed to explain the gradients.

\subsection{Efficiency of flattened and spheroidal mass distribution in explaining rotation speed fall-off}\label{sec:efficiency}

For illustration, let us compare vertical gradients for a disc-like and a spherically symmetric mass distribution with the same rotation law in the $z=0$ plane. For simplicity, we can consider the rotation law of a Kuzmin disc \citep{bib:galactic_dynamics}\[v_c(r)=\sqrt{\frac{G M}{r}}\br{1+\frac{a^2}{r^2}}^{-(3/4)}.\]
It is a simple matter to show in the quasi-circular orbit approximation (we discuss this approximation in Section \ref{sec:diskgrad}) that the vertical gradient of azimuthal velocity is
\[
-\frac{3}{2r}\sqrt{\frac{G M}{r}}
\frac{\br{a+z/r}}{\br{1+\br{a+z/r}^2}^{{7}/{4}}}
,\qquad z>0
\]
However, the absolute gradient value would be much lower for a spherically symmetric model with the same rotation curve
\[
-\frac{3}{2r}\sqrt{\frac{G M}{r}}
\frac{z/r}{\br{1+(a^2+z^2/r^2)}^{{7}/{4}}},\quad z>0.
\] and it would vanish at the disc plane. The gradients would become comparable only at some height off the disc.

In general, the contribution to the vertical gradient of rotation from a spherical potential also vanishes in the galactic mid-plane, whereas a similar contribution from the gravitational field of a disc remains non-zero. From this, it also follows that to attain the actual gradient values at larger heights off the galactic mid-plane (outside the main concentration of masses, where the 'thin disc gravitation' is comparable with the 'wide disc gravitation'), the disc model gradient must decrease in absolute value, while the spherical model gradient must increase in absolute value. Qualitatively, this explains why the vertical fall-off of rotation is, on average, weaker when the gravity of a flattened, disc-like component is dominated by the gravity of a spheroidal component. The presence of a massive spheroidal component reduces the overall vertical gradient in two ways: (i) because (as we have already seen) the contributions to vertical gradients from spheroidal components are small; (ii) because the presence of a spherical component uses up masses that would otherwise be present in the disc component, enhancing the disc-like contributions to the overall gradient.

 \subsection{Outline of the results}

In the framework of a global thin disc model approximation for the Milky Way, we use various methods for estimating the vertical gradient in rotation speed. All the methods lead to comparable results. In particular, we find the gradients for $0.22\leq\abs{z}\leq2.62\,\kpc$ by using an averaging method, mimicking realistic gradient measurements in galaxies other than the Milky Way, and for $0<\abs{z}<3.6\,\kpc$ by analysing the motion of test bodies in the gravitational potential of the Galaxy described in the thin disc model approximation. The gradient values are compared with the observed value determined by \citet{bib:gradients} from measurements of the rotation of the Milky Way in the region $\abs{z}<0.1\,\kpc$. Although our various gradient estimates are carried out at larger $z$, we consider this comparison possible. One argument for this is that the value of the velocity fall-off is almost constant with the distance off the mid-plane we encountered in our analysis. Surely, it would be best to measure the gradient values in the Galaxy at heights larger than $0.1\,\kpc$; however, such data are not available. However, we can utilize the observational fact that the gradient is constant in other galaxies. In addition, we also calculated the gradient analytically in the region $\abs{z}<0.1\,\kpc$, again obtaining results consistent with the observed gradient value in this region. The gradient values we predicted using our simple model agree amazingly well with those obtained by \cite{bib:gradients} even though the Galaxy rotation curve we utilized is unrelated to that paper. This agreement is striking in view of the fact that the gradients are usually considered to originate from more complex physics than gravitational alone, to mention only radial pressure gradients, magnetic tension, or galactic fountains.

For completeness, we compare our results with those predicted for the gradient by a customary three-component model or a maximum halo model. It turns out that the gradient features are crucially dependent on whether the gross mass distribution is more flattened or more spheroidal, and we have already encountered this feature when considering a simple example in Section \ref{sec:efficiency}.
We find that the gradient estimates for the Galaxy predicted in the disc model are in better agreement with observations than analogous predictions of models with a significant spheroidal mass component.

Finally, we apply the disc model and estimate the vertical gradients in the galaxy NGC 891, where rotation was measured far from the galactic mid-plane, again obtaining results consistent with measurements. These results show that the simple thin disc model approximation performs well in reconstructing the vertical gradient properties. Of course, this does not mean that galactic discs are thin. However, it does provide a strong argument for the fact that gross mass distribution in the Milky Way and in NGC 891 might be flattened rather than spheroidal.

\subsection{A model (SHO) rotation curve of Milky Way Galaxy}

Apart from the main reference \citep{bib:gradients}, the observational basis for our analysis is a unified set of data representing the rotation velocity of the Galaxy collected by \citeauthor{bib:sofue_unified} (2009, hereafter SHO). Based on these data, we derive a substitute thin disc surface mass density accounting for the Galaxy rotation and representing a sort of projection of Galaxy masses on to the Galactic mid-plane. If gross Galaxy matter forms a flattened structure, as seen in other spiral galaxy pictures (apart from the galactic luminous halo and the central spheroidal bulge, the latter, nevertheless, commonly described by an equivalent de Vaucouleurs disc), then the actual Galaxy gravitational potential should be better approximated by the disc model than by a model with a significant spheroidal component comprising most of the dynamical Galaxy mass inferred from the Galaxy rotation. The data set is a compilation of several independent measurements of the rotation, suitably adjusted to each other. Unfortunately, the rotation data are poor at larger distances where the measurement points are scattered randomly and are determined with large errors. This leaves too much room for arbitrariness, as neither declining nor even rising rotation could be excluded at outer radii (however, this inaccurately measured, remote Galaxy region is not very important to our analysis). Therefore, instead of the unified measurements, we decided to use a published model rotation curve that accounts for the data satisfactorily, mainly in the $\,3-8\,\kpc$ region in which we are interested. This curve was obtained by SHO for the unified measurement data by fitting a three-component model, including in addition a wavy ring pattern superposed on the exponential disc component. For brevity, we refer to this particular fit as the SHO rotation curve and denote it by $v_{\sigma}$. The SHO rotation curve agrees, as a constrained least-squares fit, with the Galaxy rotation curve almost perfectly out to large radii (however, any other fit, such as a linear combination of Hermite polynomials, a Fourier series, etc., anything one wishes, would also be acceptable, if only the fit was sufficiently smooth and close to the data in the sense of some norm, such as the least-squares method norm). The rings allowed the authors to reproduce the observed dips in rotation. Leaving the SHO mass model aside, the SHO fit alone can be treated as an actually measured rotation curve of the Galaxy, as it agrees very well with real rotation measurements in the internal Galaxy region, accounting for qualitative characteristics of the rotation. This curve agrees also with additional constraints imposed on the curve in SHO and known from separate precise single rotation measurements.\footnote{For clarity, it should be noted that a square root in the rotation law for the assumed dark halo mass distribution was unwittingly missed in SHO. This is probably not a mere misprint as a plot of the dark halo contribution to the overall rotation curve presented in their paper is parabolic at the centre, despite being linear for such a halo. Obviously, this mistake is not important to our analysis and does not affect our results.}

\section{VERTICAL VELOCITY GRADIENTS OF AZIMUTHAL VELOCITY IN THE GLOBAL THIN DISC MODEL:QUASI-CIRCULAR ORBIT APPROXIMATION}\label{sec:diskgrad}

The surface mass density $\sigma(r)$, representing a flattened mass distribution projected on to the Galactic mid-plane and corresponding to SHO rotation curve, is found in a global thin disc model approximation from
\[
\sigma(r)=
\frac{1}{\pi^2G} \vp \left[\int\limits_0^r
v_{\sigma}^2(\chi)\biggl(\frac{K\br{\frac{\chi}{r}}}{ r\
\chi}-\frac{r}{\chi}
\frac{E\br{\frac{\chi}{r}}}{r^2-\chi^2}\biggr)\ud{\chi}+\dots
\right.\nonumber\]
 \begin{equation}  \label{eq:SigmafromRotCrv} \left.
\phantom{\sigma(r)=
\frac{1}{\pi^2G} \vp (\int\limits_0^r
v_{\sigma}^2(\chi)}
\dots+  \int\limits_r^{\infty}v_{\sigma}^2(\chi)
\frac{E\br{\frac{r}{\chi}}
}{\chi^2-r^2}\,\ud{\chi}\right] \end{equation}
We derived this equation in \citet{bib:jalocha_apj}. In contrast to the equivalent relation known from the classic textbook on galactic dynamics \citep{bib:galactic_dynamics}, it does not contain derivatives of rotation. Because no additional constraints on mass distribution are taken into account in this paper, the integration in integral \eqref{eq:SigmafromRotCrv} must be cut off at some radius $R_{\sigma}$ (when integration becomes cut off, the resulting cut-off errors, as a result of some boundary effects, are different both in value and behaviour from those resulting in the classic formula). According to a criterion derived in \citet{bib:bratek_MNRAS}, the uncertainty in $\sigma(r)$ determination, resulting from the cut-off, can be neglected when $r<\frac{2}{3}\,R_{\sigma}$.
 Here, the cut-off radius equals  $R_{\sigma}=20\,\kpc$, which is the radial extent of $v_{\sigma}(r)$. In particular, the criterion is satisfied inside the ring $r\in\br{3,8}\kpc$, which is the region that interests us.\footnote{
  The lower bound for the Galaxy mass in the global disc model, integrated out to cut-off radius $R_{\sigma}$, is $1.12\times10^{11}\msun$  for the SHO rotation curve, using a substitute surface density defined in equation  \eqref{eq:SigmafromRotCrv}.}

Given a surface mass density, azimuthal velocities of test particles in the gravitational field of a thin disc can be estimated based on the radial force equilibrium condition \begin{equation}\label{eq:condition}\frac{\,v_{\varphi}^2(r,z)}{r}\approx -g_r(r,z),
 \end{equation} Here, $g_r(r,z)$ is the radial component of the gravitational acceleration (this approximation is justified later), and hence
\[\frac{v_{\varphi}^2(r,z)}{r}= {G \int\limits_0^\infty\!\!\int\limits_{0}^{2\pi}\frac{ \br{r-\chi
\cos{\varphi}}\sigma\br{\chi}\chi\ud{\chi}\ud{\varphi}}{\br{r^{2}+\chi^{2}-2  r \chi
\cos{\varphi}+z^{2}}^{3/2}}}\,.
\]
Integration with respect to the azimuthal angle gives
\[
{v_{\varphi}^2(r,z)}=\!\int\limits_0^{\infty}\!\!\!\frac{2\, G\sigma\br{\chi{}}\chi{}\ud{\chi{}}}{\sqrt{\br{r-\chi{}}^2+z^2}}\!\br{\!K\sq{X}-
\frac{\chi^2-r^2+z^2}{\br{r+\chi}^2+z^2}\,E\sq{X}\!}\!,\]
\begin{equation}\label{eq:VOverDiskFromSigma} X=-\sqrt{\frac{4r\chi{}}{\br{r-\chi{}}^2+z^2}}<0.\end{equation}
As an aside, we remark that the velocity field ${v_{\varphi}(r,z)}$ could be equally well calculated without the intermediate step of determining $\sigma(r)$. In this respect, we should first find the kernel function
 \[K(r,z;\chi)=\int\limits_{0}^{\infty}\lambda J_1\br{\lambda\chi}J_1\br{\lambda{r}}\exp\br{-\lambda\abs{z}}\ud{\lambda}, \quad\chi,\,r>0.\]
 In particular, in the limit $z\to0$, $K(r,z;\chi)$ can be expressed by the Dirac $\delta$-function,  $K(r,0;\chi)=r^{-1}\delta(r-\chi)=\br{r\chi}^{-1/2}\delta(r-\chi)$.
With the aid of this kernel, we can show, on combining equations (A1) and (A5) from the appendix in \citet{bib:bratek_MNRAS} and using equation \eqref{eq:condition}, that the following concise relation between the velocity field $v_{\varphi}(r,z)$ and the rotation curve $v_{\varphi}(r,0)\equiv{}v_{\sigma}(r)$ holds:    \begin{equation}\label{eq:VFromKernel}
{v_{\varphi}^2(r,z)}=r\int\limits_0^{\infty}K(r,z;\chi)v^2_{\sigma}(\chi)\ud{\chi}.\end{equation}
Obviously, condition \eqref{eq:condition}, and thus equations \eqref{eq:VOverDiskFromSigma} and \eqref{eq:VFromKernel}, hold for azimuthal velocity only approximately, as orbits of test bodies form very complicated spatial curves, occasionally crossing the galactic mid-plane. These equations, however, should hold approximately for orbits whose projections on to the galactic mid-plane are almost circles. For such orbits, equations \eqref{eq:VOverDiskFromSigma} or \eqref{eq:VFromKernel} could be used for estimating the vertical gradients of azimuthal velocity by direct calculation of $\partial_zv_{\varphi}(r,z)$.
Before proceeding further, we first try to convince ourselves that the above approximation can indeed be applied.

By taking into account the first integral of motion existing by axial symmetry, $r^2(t){\varphi'(t)}=j_z$, where $j_z$ s a constant characterizing a given trajectory, the radial component of the equations of motion $r''(t)-r(t)\br{\varphi'(t)}^2=-\partial_r\Phi(r(t),z(t))$ can be recast in the form of an equation for the image of the trajectory projected on to the galactic mid-plane:
\begin{equation}\label{eq:trajectory}\frac{r^2(\varphi)+2r'(\varphi)^2-r(\varphi)r''(\varphi)}{r^2(\varphi)}=
-\frac{
r(\varphi)g_r(r(\varphi),z(\varphi))}{v_{\varphi}^2(r(\varphi),z(\varphi))},\end{equation}
Here, $v_{\varphi}(r(\varphi),z(\varphi))=r(\varphi)\varphi'(t(\varphi))=j_z/r(\varphi)$ and $g_r(r(\varphi),z(\varphi))=-\partial_{r(\varphi)}\Phi(r(\varphi),z(\varphi))$ is the gravitational acceleration experienced on the trajectory. The term on the left-hand side of equation \eqref{eq:trajectory} is proportional to the curvature $\kappa_{xy}(\varphi)$ of the projection \[\kappa_{xy}(\varphi)=
\frac{
\abs{r^2(\varphi)+2r'^2(\varphi)-r(\varphi)r''(\varphi)}}{\br{r^2(\varphi)+r'^2(\varphi)}^{3/2}}.\] Thus, equation
\eqref{eq:VOverDiskFromSigma} is exactly satisfied for orbits for which the left-hand side of equation \eqref{eq:trajectory} identically equals $1$. This is true for a trajectory whose projection on to the galactic plane forms a circle, $r(\varphi)=const.$, or a hyperbolic spiral,
 $r(\varphi)=a\br{\varphi-\varphi_o}^{-1}$. As for realistic trajectories, equation (5) should work in the approximation of quasi-circular orbits. A quasi-circular orbit by definition can be enclosed within a relatively thin toroidal tube symmetric with respect to the galactic mid-plane and concentric with the centre. The projection of such orbits on to the galactic mid-plane is roughly circular.

\begin{figure}
    \centering
         \includegraphics[width=0.5\textwidth]{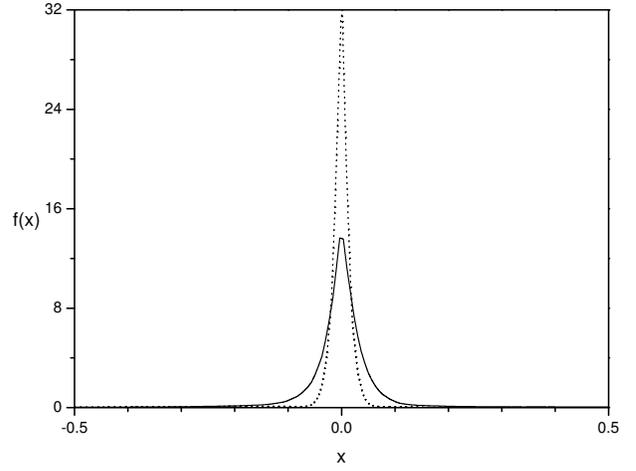}
         \vspace{-0.02\textheight}
    \caption{\label{fig:distrib} Distribution functions for the radial force equilibrium test (solid line) and for the orbit circularity test (dotted line); these smooth-looking lines were obtained by filtering data using averaging in a window of small x-width. The dotted line shows the superposition of distribution functions of variable  $x=\log_{10}\frac{r}{\bar{r}}$   measuring departure of radial distance $r$ from its average value $\bar{r}$ on a given orbit, obtained for various trajectories (here, $\bar{x}=0.03$ and $\delta{x}=0.3$). The solid line shows the analogously obtained distribution function for variable $x=\log_{10}{v^2_{\varphi}}/\br{r g_r}$, (here, $\bar{x}=-0.014$ $\delta{x}=0.09$)  measuring the departure of 'centrifugal force` ${v^2_{\varphi}}/{r}$ from the actual magnitude of the radial component of the gravitational acceleration.}
  \end{figure}

 The expectation that equation \eqref{eq:VOverDiskFromSigma} holds approximately for quasi-circular orbits has been confirmed by the following numerical experiment. Starting from various initial conditions, trajectories of test bodies were found by numerical integration of equations of motion in the gravitational field of the disc of surface density \eqref{eq:SigmafromRotCrv} corresponding to the SHO rotation curve. The initial positions were generated randomly in space with probability density proportional to the surface mass density and falling off exponentially with $z$. At the initial instant of each simulation, the only non-zero velocity component assumed non-zero was the azimuthal one. For a given initial position, the initial velocity was calculated from equation \eqref{eq:VOverDiskFromSigma}.
    To test our hypothesis, we found nearly 300 trajectories with various complicated shapes. For the majority of these, the radial variable on a given orbit was observed to have small dispersion relative to the mean radial variable. The quasi-circular orbit condition \eqref{eq:condition}  was also satisfied with small dispersion. The relevant statistical analysis is presented and explained in more detail in Fig. \ref{fig:distrib}.
From the point of view of the precision of astronomical data and the accuracy of rotation curve modelling, such a result can be regarded at least as satisfactory. This way, the approximate relation \eqref{eq:VOverDiskFromSigma} has been justified.

For our needs, approximation $\eqref{eq:VOverDiskFromSigma}$ is also justified by the fact that the results for gradient estimates in \citet{bib:gradients}, which we aim to reconstruct in the global disc model approximation, were obtained under the idealized assumption of non-intersecting gas orbits with vanishing velocity in the $z$-direction (including orbits off the galactic mid-plane). This is also the central assumption in the method of deriving the azimuthal velocity profile as a function of the radial distance from the Galactic Centre, at least for the part of rotation curve of interest in \citet{bib:gradients}.
Our approximation is thus sufficient, as it is no less accurate than the method of obtaining the rotation curve on different heights off the Galactic mid-plane. Hence, there is no reason or need to make our equations more realistic than they are in the present form. All in all, what we indeed assume in our paper, which may be considered non-standard, is that gross dynamical Galaxy mass forms a flattened rather than spheroidal object.

\subsection{Analytical estimates of vertical gradients in the global thin disc model}\label{sec:diskgrad2}

\begin{figure}
    \centering
       \includegraphics[width=0.5\textwidth]{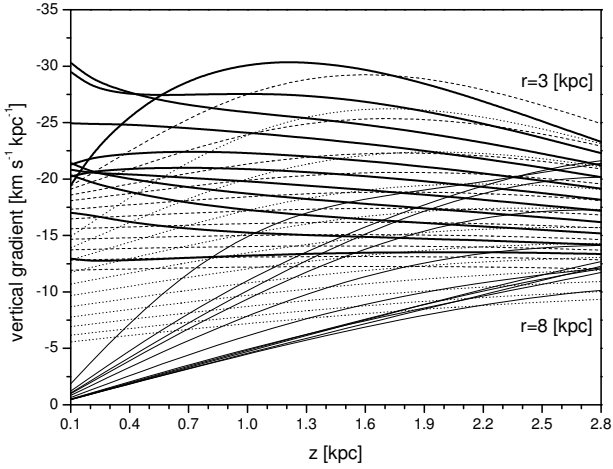}
       \caption{\label{fig:gradient_analytical_disk_model}  Vertical gradient of azimuthal velocity for quasi-circular orbits calculated in a disc model (thick line), in a three-component model with a light dark halo (dashed line), in a three-component model with a massive dark halo (dotted line) and in a maximal halo model (thin line), all discussed in the text, shown for different values of radial variable $r\in\br{3,8}\kpc$ in steps of $\Delta r=0.5\,\kpc$. }
  \end{figure}

 The vertical gradient of azimuthal velocity for quasi-circular orbits can be found directly from equation \eqref{eq:VOverDiskFromSigma} by differentiation with respect to $z$:
\begin{eqnarray}\label{eq:diskgradient}\partial_z{}v_{\varphi}\br{r,z}=\frac{Gz}{v_{\varphi}(r,z)}\int\limits_0^{\infty} \frac{\chi \sigma (\chi )\ud{\chi}}{{\left( z^2 + {\left( r - \chi  \right) }^2 \right) }^{\frac{3}{2}}}\times\cdots \end{eqnarray}
   \[
\left(\!\frac{7r^4 \!+\! 6r^2\!\left(\! z^2\! -\! {\chi }^2 \right)\!  -\! {\left(
\! z^2\! +\! {\chi }^2 \right)\! }^2}
  {{\left( z^2 + {\left( r + \chi  \right) }^2 \right) }^2}E\sq{-\sqrt{\frac{4r\chi{}}{\br{r-\chi{}}^2+z^2}}}
  +\cdots\right.\]
  \[\phantom{xxxxxXX}
  \left.\cdots+
  \frac{-r^2 + z^2 + {\chi }^2}{z^2 + {\left( r + \chi  \right) }^2}K\br{-\sqrt{\frac{4r\chi{}}{\br{r-\chi{}}^2+z^2}}}        \right).
\]
Various gradient lines are depicted in Fig. \ref{fig:gradient_analytical_disk_model} (including the analogous lines of other models studied in the following sections).
  It is seen from this figure that the absolute gradient value decreases with the growing mass of the spherical mass component. It is also evident that the disc model is naturally suited to describe large vertical gradients of rotation observed close to the galactic mid-plane. A significant spherical component reduces the gradient value in two ways: it has a necessarily small contribution to the resultant gradient close to the galactic mid-plane; it 'removes' mass from the disc component, reducing in turn the disc contribution to the overall gradient. The fact that the disc component gives a larger contribution than the spherical component to the vertical gradient in the vicinity of the galactic mid-plane is well illustrated in the example of the Kuzmin disc discussed in Section \ref{sec:summary}.

\begin{figure}
    \centering
       \includegraphics[width=0.5\textwidth]{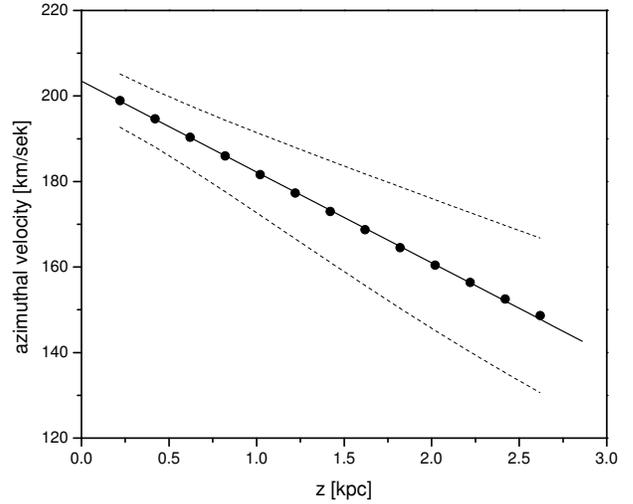}
       \caption{\label{fig:gradient_disk_model} Variation of azimuthal velocity on quasi-circular orbits as a function of distance off the Galactic mid-plane in the global disc model. Each point represents the mean azimuthal velocity component over the radial interval $r\in\br{3,8}\kpc$ at a given height off the Galactic mid-plane, calculated using equation \eqref{eq:VOverDiskFromSigma}. The dashed lines represent the corresponding one-standard deviation band. The linear regression fit slope is $-21.3\pm4.2\,\gu$.}
      \end{figure}

  In what follows, we determine a global vertical gradient in the Galaxy by averaging the behaviour of local rotation fall-offs in the region $r\in(0,8)\,\kpc$ in a way that mimicks gradient measurements in galaxies. This enables a comparison with estimates of vertical derivatives of the Galaxy rotation curve reported by \citet{bib:gradients}.
  To this end, we first prepared an array of azimuthal velocities calculated from equation \eqref{eq:VOverDiskFromSigma} for pairs $(r,z)$, $r\in\br{3,8}\,\kpc$ and $z\in\br{0.22,2.62}\,\kpc$,  in steps of $\Delta{}r=1\,\kpc$ and $\Delta{}z=0.2\,\kpc$. This assumes that the weighting function is homogeneous in space. Next, for a given $z$, we calculated the mean azimuthal velocity in the radial interval and the corresponding standard deviation from the mean in this interval. The mean velocity is a reasonable estimate, as the Galaxy rotation curve is roughly flat over this interval. The global vertical gradient and its error can be now determined by finding the slope of a linear regression fit to these data (see Fig. \ref{fig:gradient_disk_model}).    

  In this way, we find that the vertical gradient of azimuthal velocity for quasi-circular orbits in the global disc model approximation is
 $-21.3\pm4.2\,\gu$,
whereas the Galaxy gradient measurements yield $-22\pm6$ \citep{bib:gradients}.

  The preceding analysis has assumed that the variable $z$ interval is greater than and lies outside the measurement interval  $\abs{z}<0.1\,\kpc$ used in \citep{bib:gradients}. However, because of the linearity of the fall-off of the azimuthal velocity evident in Fig. \ref{fig:gradient_disk_model}, the disc model gradient value can be extrapolated in the direct vicinity of the galactic mid-plane, where the actual gradient should be comparable (the disc model might give absolute values of the gradient at the galactic mid-plane greater than those observed, but this would only be the model effect caused by the assumed infinitely thin layer of disc mass). For more realistic finite width discs, the absolute gradient value would be lower at the Galactic mid-plane, but it would still remain slightly greater than or comparable with the gradient values at larger heights off the plane, larger than the stellar disc width (of the order of $0.3\kpc$), where both the thin layer disc and a finite width disc must surely give comparable predictions. As seen in Fig. \ref{fig:gradient_disk_model}, the thin disc gradients are almost constant at heights several times greater than the width of the Galaxy stellar disc. We stress the observational fact that the absolute gradient values off the Galactic mid-plane, which we predict for the Galaxy, are consistent with the fall-off measured in haloes of other galaxies (compare the discussion in the summary of \citep{bib:gradients}), This suggests that the gradients are indeed constant at large distances off Galactic mid-planes. Because for remote galaxies such measurements must necessarily cover a broader $z$ interval, as we have assumed, this gives observational support for the fact that the gradient should be almost constant close to the Galactic disc.

\begin{figure}
    \centering
       \includegraphics[width=0.5\textwidth]{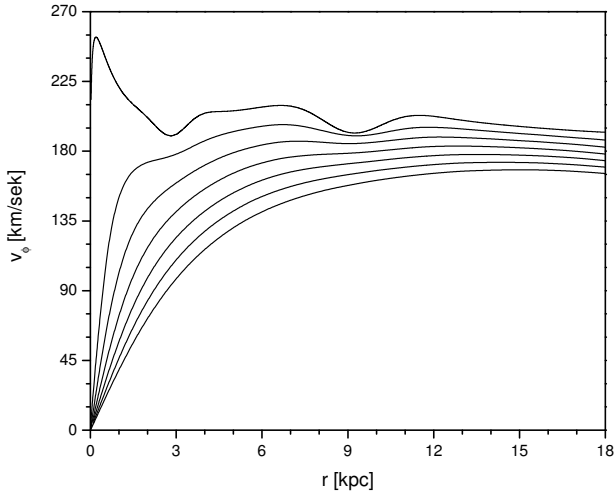}
       \caption{\label{fig:predkosci_dysk}
        Azimuthal velocity curves, $v_{\varphi}(r,z)$, obtained in the global disc model approximation for quasi-circular orbits as a function of distance from the Galactic mid-plane, shown in steps of $\Delta z=0.6\kpc$. The absolute value of the vertical gradient of azimuthal velocity grows with the separation $\Delta v_{\varphi}$ of the curves in this diagram.}
  \end{figure}

For brevity, we call the above method of global gradient determination the 'I-method'.
 The I-method works well when calculations are carried out over the flat part of the rotation curve, otherwise the velocity dispersion might be large, leading to large uncertainties in the gradient determination. However, the global gradient in the region of interest can also be estimated by finding a mean value over a radial interval of local vertical gradients determined at each $r$ separately. By a local gradient at a given $r$ we mean a slope of a linear regression fit to the values of azimuthal velocities calculated from equation \eqref{eq:VOverDiskFromSigma}, by assuming various $z$ in the same region as in the I-method. We call this method of global gradient determination the 'II-method', to distinguish it from the I-method. A global gradient obtained thus is $-21.08\pm 5.75\, \gu$.
   The standard deviation of this value, relative to the mean value, measures the degree of change of the gradient with the radial variable. Fig.   \ref{fig:predkosci_dysk} illustrates this change by showing rotation curves at different heights off the Galactic mid-plane.

  Finally, it is worth noting that the vertical fall-off of the rotation curve determined by a linear fit
  \[v_{\varphi}(r,z)=v_{\sigma}(r)+\gamma z+\delta\abs{z},\qquad \delta=0\]
  to the data array considered in this section is $\gamma=21.2\pm2.4\,\gu$ with $95$ per cent confidence level, again consistent with the measurements. We have omitted the rolling term $\delta |z|$, as the disc model assumes $z$-reflection symmetry. This method is analogous to that used in \citep{bib:gradients} for the determination of the gradient value from Galaxy rotation measurements.

  \subsubsection{Vertical gradient of rotation close to the Galactic mid-plane (at $\abs{z}<0.1\,\kpc$)}
\begin{figure}
    \centering
       \includegraphics[width=0.5\textwidth]{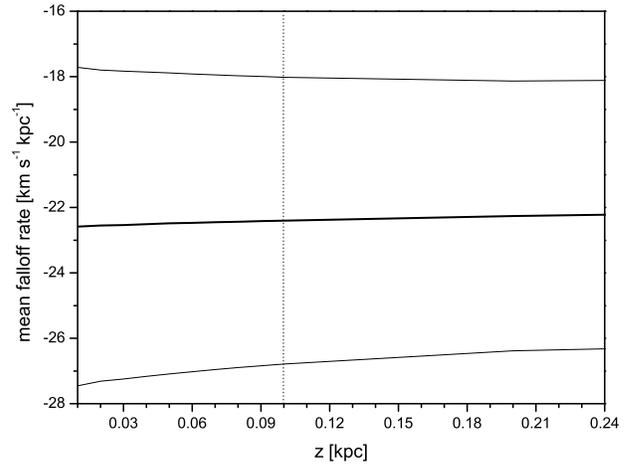}
       \caption{\label{fig:mean_gradient}
       Behaviour of the vertical gradient in Galaxy rotation speed at small heights off the Galactic mid-plane. The thick line shows the mean gradient value at a given $z$ averaged over interval $r\in\br{3,8}\,\kpc$, calculated with the aid of equation \eqref{eq:diskgradient}. The thin lines show the one-standard deviation from the mean value.
       }
  \end{figure}

  Above, we have estimated the vertical gradient for the Galaxy using the rotation speed (equation \eqref{eq:VOverDiskFromSigma}) outside the strip $\abs{z}<0.1\,\kpc$. This choice was dictated by the desire to avoid difficulties in numerical integration close to the mid-plane where integrands in equation \eqref{eq:diskgradient} become divergent at $z=0$ and thus numerically intractable close to $z=0$.
   We have also given some arguments for the fact that our gradient estimates could be extrapolated toward the close vicinity of the mid-plane $\abs{z}<0.1\,\kpc$ and thus could also be compared with the value determined from observations in this region. However, to substantiate our statements, we independently attempted to calculate the gradient in the region $\abs{z}<0.1\,\kpc$ directly from integral \eqref{eq:diskgradient}, by carrying out regularized integration in the principal value sense. Numerically, this could be achieved by applying the integration rule
 \[ \int\limits_0^{R}f(r,\chi)\,\ud{\chi}\to
  \int\limits_0^{r(1-\varepsilon)}f(r,\chi)\,\ud{\chi}+
\int\limits_{r(1+\varepsilon)}^Rf(r,\chi)\,\ud{\chi},\] with $\varepsilon$ being some tiny number, in practice of the order of $10^{-6}$.
 As seen from Fig. \ref{fig:mean_gradient}, on the lower height scales $\abs{z}<0.1\,\kpc$ the predicted gradient value does not change noticeably. It smoothly overlaps with the upper region value and is still consistent with the observed gradient value. 
          This is also the upper bound on the absolute gradient value expected from other mass models; in particular, a finite disc absolute gradient value should be comparable, provided the disc comprises gross dynamical mass.

  \subsection{Vertical gradient of rotation from motions of test bodies}
  The most convincing estimation of the vertical gradient in the framework of the global disc model seems to be the statistical analysis of trajectories of test bodies. The trajectories are found by numerical integration of the equations of motion in the Galaxy gravitational field produced by the disc. The vertical velocity gradients are estimated by the slope of a linear regression fit on the plane of values $(z,v_{\varphi})$, typical of each trajectory separately.

  Given a trajectory, we can determine its different characteristics such as the average radial distance, average azimuthal velocity, average absolute distance off the Galactic mid-plane, the $z$-variable dispersion, etc. In our context, it seems most appropriate to use temporal averaging rather than any other averaging with respect to the azimuthal angle. \footnote{Let $\varphi(t)$ describe the time dependence of the angular position on a trajectory, $\Delta\varphi\equiv\varphi(T)-\varphi(0)$, and let $T$ be the averaging time. A temporal average of a function $u(\varphi(t))$ along the trajectory is defined as $\langle{u}\rangle_T=\frac{1}{T}\int_{T} u(\varphi(t))\ud{t}$ and, similarly, an angular average is $\langle{u}\rangle_{\Delta{\varphi}}=\frac{1}{\Delta{\varphi}}\int_{\Delta{\varphi}}u(\varphi)\ud{\varphi}$.
  On changing variables, the two averages can be compared with each other. Noting that $r(\varphi(t)){\varphi}'(t)=v_{\varphi}(t)$, it follows that  $\langle{u}\rangle_{\Delta{\varphi}}=\frac{1}{\Delta{\varphi}}
  \int_{T}u(\varphi(t))\frac{v_{\varphi}(t)}{r(\varphi(t))}\ud{t}\neq\langle{u}\rangle_{T}$. Thus, $\langle{u}\rangle_{\Delta{\varphi}}\neq\langle{u}\rangle_{T}$, unless $\varphi'(t)=\mathrm{const}.$}
    For nearly circular orbits in the Galactic plane, these averages would be almost the same; however, the results would differ from each other for more complicated motions. The problem of choosing appropriate averaging methods is general, and most difficult to solve in data analysis. Statistical analysis as such is clear, but the real problem is to define the appropriate space of events and the probability density defined on it. In this respect, symmetry arguments or physics hidden behind a particular problem may help.

    By $z$-reflection symmetry, it is expected that variable $z$ on average, should be zero along a given trajectory. The mean absolute distance from the Galactic mid-plane, or standard deviation of variable $z$,
   can be regarded as the simplest measures of a typical distance from the mid-plane on a given trajectory. Unfortunately, the notion of the typical distance is a matter of convention. For example, for a $z$-symmetric Gaussian density function, the two 'typical distances' are comparable; however, the standard deviation of variable $z$ from $z=0$ is about $1.25$ times greater than the mean absolute distance from $z=0$.
    Another problem is the choice of a weighting function. The assumption of the temporal averaging method is tantamount to the statement that the weight is uniform in time, and each instant of time is treated on the same footing. In this case, more probable are events that, on average, last longer. For a stationary system (which has energy as a constant of motion), the assumption of temporal averaging is thus justified. However, in the disc approximation also, angular averaging is justified because of the assumed axial symmetry. In the latter case, the weighting function should be taken as homogeneous in the angular variable. Thus, symmetry arguments cannot differentiate between two possible weighting functions, at least for the totality of all trajectories satisfying various initial conditions distributed axially symmetric. Fortunately, the averaging method and the choice of distance measure do not drastically influence the predicted vertical gradient values, as far as the accuracy of galaxy modelling is concerned.

\begin{figure}
  \vspace{-0.6cm}
    \centering
       \includegraphics[width=0.45\textwidth]{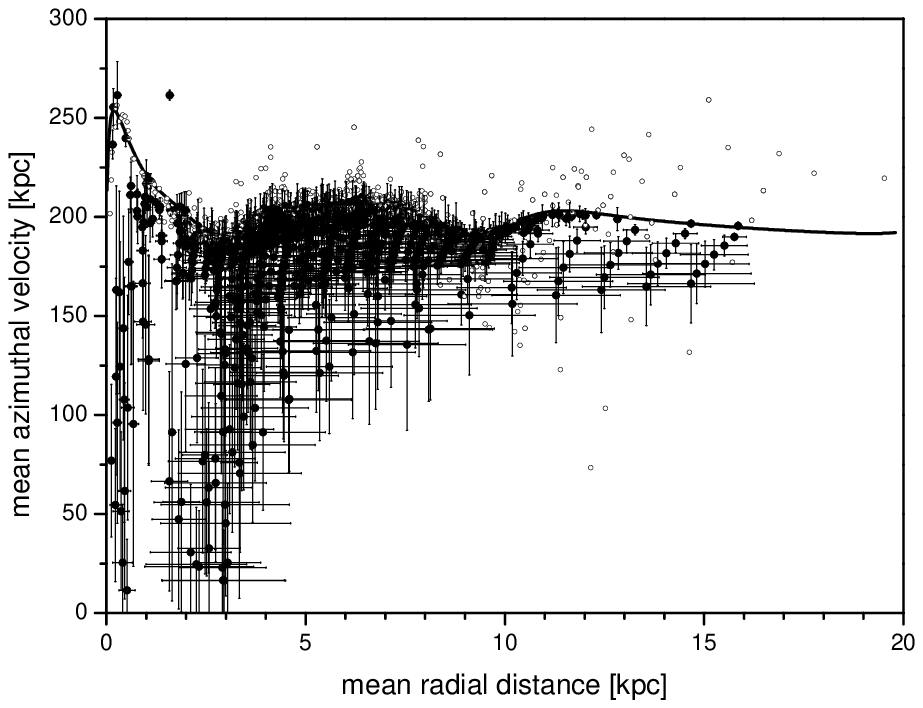}
       \includegraphics[width=0.45\textwidth]{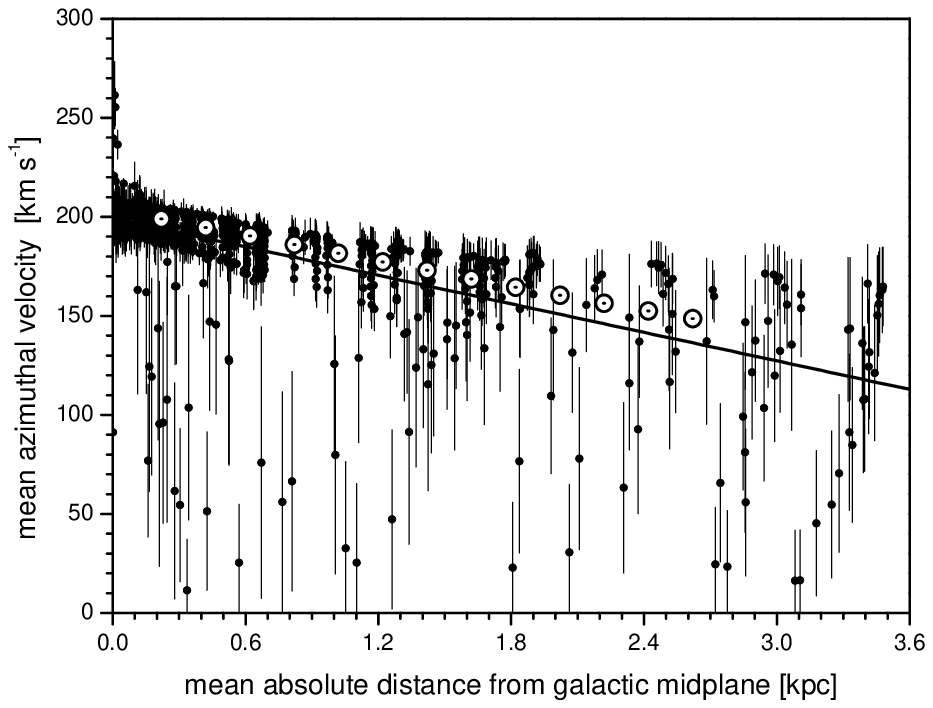}
       \includegraphics[width=0.45\textwidth]{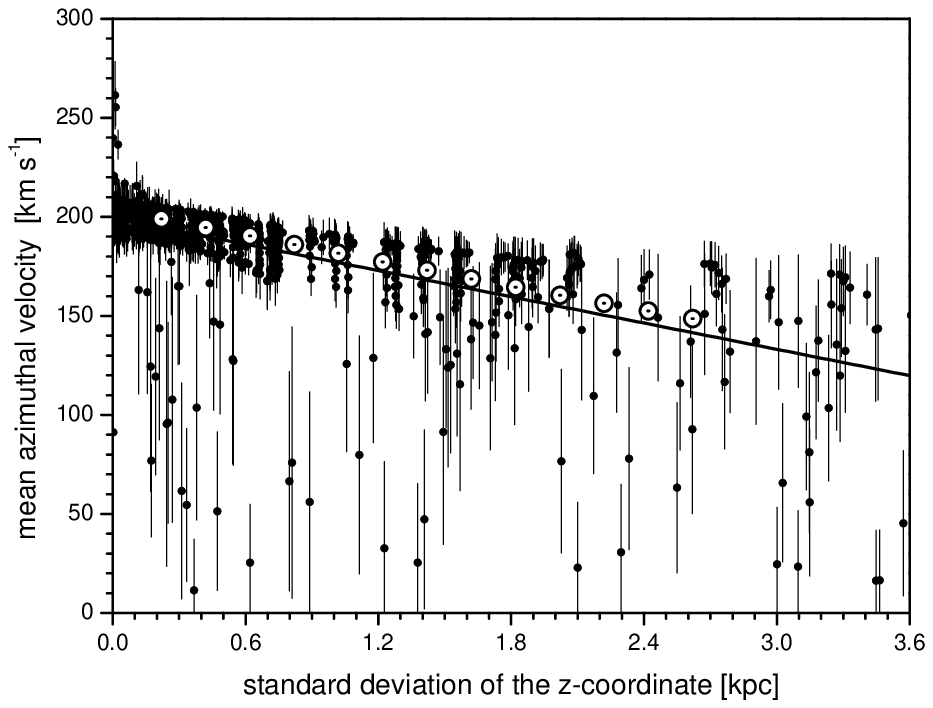}
       \caption{\label{fig:simulation}
     Results of numerical simulations obtained for trajectories of test bodies moving in the gravitational field found in the disc model approximation based on the SHO Galaxy rotation curve. The top figure shows the mean azimuthal velocity (solid circles) as a function of the mean radial distance obtained for various trajectories (standard deviations from these quantities are represented by vertical and horizontal segments), shown together with the real rotation curve measurements (empty circles) and with the SHO model rotation curve (solid curve). In the central and bottom figures, solid circles represent the mean values of the azimuthal velocity, and the respective standard deviations are represented by vertical segments; both quantities were calculated for each trajectory separately. These averages are shown as a function of the mean absolute distance from the Galactic mid-plane (central figure) or as a function of the standard deviation of the $z$-variable (the mean $z$-variable approximately equals zero; bottom figure). To obtain these data, temporal weighting was used. The solid lines are the linear regression fits to these data, with slopes $-24.0\pm2.1$ and $-22.1\pm1.9$ (with $95$ per cent confidence level) in the central and bottom figures, respectively. For comparison with the slopes, the empty circles represent the data copied from Fig. \ref{fig:gradient_disk_model}. }
  \end{figure}

  The results of our analysis are presented and explained in Fig. \ref{fig:simulation}.
By assuming the mean absolute distance from the Galactic mid-plane as the typical distance on a trajectory, we obtain for the gradient values the following estimates, $-24.0\pm2.1\, \gu$ and $-17.8\pm1.1\, \gu$ with $95$ per cent confidence, respectively, for temporal averaging and angular averaging along particular trajectories. To carry out the estimates, we took into account about $600$ trajectories with equally spaced initial positions and about $300$ additional trajectories already used in Section \ref{sec:diskgrad} with the initial positions chosen at random with probability proportional to the disc mass density. As we have already noted, the gradient estimates will change depending on what typical distance off the mid-plane is chosen. For example, the absolute values of the gradient should be expected to be lower for the maximum distance as the typical distance. Undoubtedly, the most reasonable measure of the typical distance is simply the dispersion of the $z$ variable on a trajectory. Then, linear fits to temporally averaged data on each trajectory give $-22.1\pm1.9\, \gu$ with $95$ per cent confidence level, whereas the angular-averaging give $-16.3\pm1.0\, \gu$ for an analogous fit and with the same confidence level. As expected, the various gradient values are scaling with the assumed typical distance. For example, with temporal averaging, the ratio of $\langle\abs{z}\rangle_T$ to $
  \sqrt{\langle\br{z-\langle{z}\rangle_T}^2\rangle_T}$ is $0.9146\pm0.0007$, which is comparable with $0.92$ -- the inverse of the respective gradient values.
\medskip

   Concluding Section \ref{sec:diskgrad}, we note that the various, independent methods for the vertical gradient estimates presented in this section give results consistent with each other. Also, they all agree with the observed global fall-off of rotation from the Galactic mid-plane of $-22\pm 6\, \gu$ determined by \citet{bib:gradients}. In particular, this value best agrees with the slope $-22.1\pm1.9\, \gu$ of our linear regression fit to a sequence of pairs  $\{{\sqrt{\langle\br{z-\langle{z}\rangle_T}^2\rangle_T}, \langle{v_{\varphi}}\rangle_T}\}$  for the sequence of the examined trajectories of test bodies.

\section{COMPARISON WITH VERTICAL GRADIENT VALUES IN OTHER MODELS OF THE GALAXY }\label{sec:customary}

It is interesting to see how the results of Section \ref{sec:diskgrad} compare with those of other, customary simple Galaxy models. For this purpose, we first consider a three-component Galaxy model consisting of a central spherically symmetric bulge, an exponential thin disc and a spherical dark halo. Next, we consider a (spherically symmetric) maximal halo model.

\subsection{Vertical gradients in a three-component Galaxy model}

The central bulge volume density $\rho_b(r)$ is obtained in a standard way, by deprojection of the corresponding surface (or column) mass density  $\sigma_b(r)$
with the help of the following Abel relation:
\[\rho_b(r)=-\frac{1}{\pi}\int\limits_{r}^{\infty}
\frac{\sigma_b(\xi)\ud{\xi}}{\sqrt{\xi^2-r^2}}.\]
 The surface density is assumed proportional to the empirical brightness law of  \citet{bib:Vaucouleurs}  which assumes a constant mass-to-light ratio in the bulge component:
\[\sigma_b(r)=\Xi_o\exp\br{-\kappa\br{\chi}^{1/4}},\qquad \chi=\frac{r\kappa^4}{R_o}.\]
 The corresponding circular speed in the disc plane can be written in analytical form \footnote{$\mathcal{M}(x)$ is expressed by Meijer G-function \citep{bib:Ryzhik}   \\ $\!\!\mathcal{M}(x)=G^{8,1}_{1,9}\br{x\left|\begin{array}{c}- \frac{1}{8}\\
   0,\frac{1}{8},\frac{1}{4},\frac{3}{8},\frac{3}{8},\frac{1}{2},
  \frac{5}{8},\frac{3}{4},- \frac{9}{8}
   \end{array}\right.    }$ }
\[v_{0b}\br{r}=\sqrt{ \frac{e^{\kappa}GR_0\Xi_0}{16\pi^3\kappa^4}\chi^{5/4}
\mathcal{M}\br{\frac{\chi^2}{4096^2}}},\qquad \chi=\frac{r\kappa^4}{R_o}.\]
The galactic disc is assumed to be exponential
\[\sigma_d(r)=\Theta\exp\br{-2\chi},\qquad \chi=\frac{r}{2\epsilon_o}.\] Finding of the velocity on circular orbits in the disc plane is a textbook problem with solution
\[v_{0d}(r)=\sqrt{ 4\pi G \epsilon_o\Theta\chi^2\br{I_0(\chi)K_0(\chi)-I_1(\chi)K_1(\chi)}},\]
 \[\chi=\frac{r}{2\epsilon_o}.\]The third component is a spherically symmetric (dark) halo with volume density \[\rho_h(r,z)=\frac{a_ob_o^2}{b_o^2+{{r^2+z^2}}}.\] The rotational velocity on circular orbits is also a textbook result:  \[v_{0h}(r)=\sqrt{4\pi G a_o b_o^2\br{1-\frac{b_o}{r}\arctan\br{\frac{r}{b_o}}}},\qquad z=0.\] Hence, the model rotation curve in the galactic mid-plane is $v_c(r)=\sqrt{v_{0b}^2+v_{0d}^2+v_{0h}^2}$.
Outside this plane, for $z\ne0$, the radial component $g_r(r,z)$ of the gravitational acceleration can be still found. The azimuthal component of velocity on quasi-circular orbits in this field is estimated in the same way as in Section \ref{sec:diskgrad} from the approximated formula $v_{\varphi}^2\br{r,z}=-rg_r(r,z)$ which has there been shown to be satisfied in a statistical sense. Again, we come to the conclusion that the azimuthal velocity can be decomposed such that
$v_{\varphi}(r,z)=\sqrt{v_b^2(r,z)+v_d^2(r,z)+v_h^2(r,z)}$, where
\[v_b(r,z)=\frac{\kappa^4 r}{R_o\chi}v_{0b}\br{\chi},\qquad \chi=\frac{\kappa^4\sqrt{r^2+z^2}}{{R_o}}\]
\[v_h(r,z)=\frac{r}{\sqrt{r^2+z^2}}v_{0h}\br{\sqrt{r^2+z^2}},\]
The azimuthal velocity for the disc component $v_d(r,z)$ has already been given in equation \eqref{eq:VOverDiskFromSigma}, in which we should put $\sigma(r)=\sigma_d(r)=\Theta\exp\br{-r/\epsilon_o}$. Next, we calculate the vertical gradient of azimuthal velocity:
\begin{equation}\label{eq:3compgradient}\partial_zv_{\varphi}=\frac{v_b\partial_zv_b+v_d\partial_zv_d+
v_h\partial_zv_h}{
\sqrt{v_b^2+v_d^2+v_h^2}}.\end{equation} The calculation of gradients for $v_b$ and $v_h$ is easy. For $v_d$  it has already been given in equation \eqref{eq:diskgradient}, where again we put  $\sigma(r)=\sigma_d(r)$.

\begin{figure}
    \centering
       \includegraphics[width=0.5\textwidth]{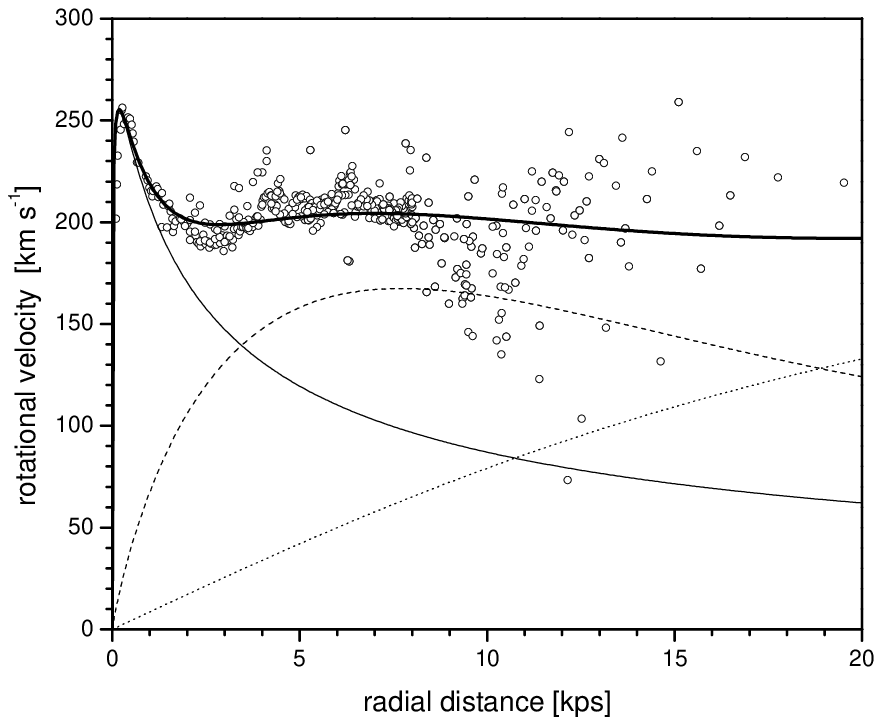}
        \includegraphics[width=0.5\textwidth]{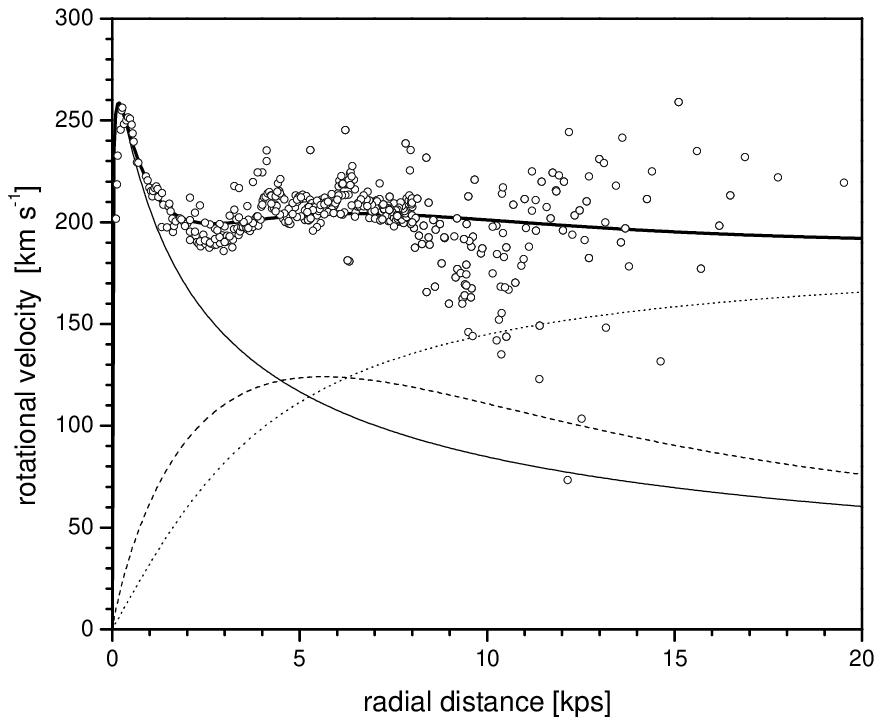}
       \caption{\label{fig:3model}Two least-squares fits of a three-component Galaxy model considered in the text fitted to the SHO rotation curve (solid lines), with the smaller dark halo (top figure) and with the larger dark halo (bottom figure), shown together with the unified rotation velocity measurements collected by SHO (empty circles). The model rotation curves are decomposed to the central bulge (thin line), the exponential disc (dashed line) and to the dark halo (dotted line).
         }
  \end{figure}

In the case of the three-component model, we examined two example fits that accounted for the Galaxy rotation with various dark halo masses. The masses of the bulge, disc and dark halo are $1.79\times10^{10}\msun$, $5.90 \times10^{10}\msun$ and  $8.21 \times10^{10}\msun$ (total $1.59\times10^{11}\msun$) for the model with smaller halo,(
   $\kappa  = 7.67,\quad
   {Ro} = 0.606\,\kpc,\quad
      {\Xi o} = 2.16\times{10}^3\,\msun\pc^{-2},\quad
     \Theta  = 7.45\times{10}^2\,\msun\pc^{-2},\quad
   {\epsilon o} = 3.6\,\kpc, \quad
     {ao} = 4.1\times 10^{-3}\,\msun\pc^{-3},\quad
   {bo} = 18\,\kpc,\quad{R_{max}} = 20\,\kpc$)
   and  $1.70 \times10^{10}\msun$,\quad $2.41\times 10^{10}\msun$ and $1.28\times 10^{11}\msun$ (total $1.69\times 10^{11}\msun$) for the model with  larger halo.(
   $\kappa  = 7.67,\quad{Ro} = 0.558\,\kpc,\quad
   {\Xi o} = 2.41\times{10}^3\,\msun\pc^{-2},\quad
    \Theta  = 5.62\times{10}^2\,\msun\pc^{-2},\quad
   {\epsilon o} = 2.62\,\kpc,$ $\quad
     {ao} = 60.6\times10^{-3}\,\msun\pc^{-3},\quad
   {bo} = 3.31\,\kpc,\quad{R_{max}} = 20\,\kpc$)
The rotation curves of the two models are shown in Fig. \ref{fig:3model}.
The behaviour of the vertical gradient (equation \eqref{eq:3compgradient}) of the azimuthal velocity as a function of the distance off the Galactic mid-plane is shown for various $r$ in Fig. \ref{fig:gradient_analytical_disk_model}.

\begin{figure}
    \centering
       \includegraphics[width=0.5\textwidth]{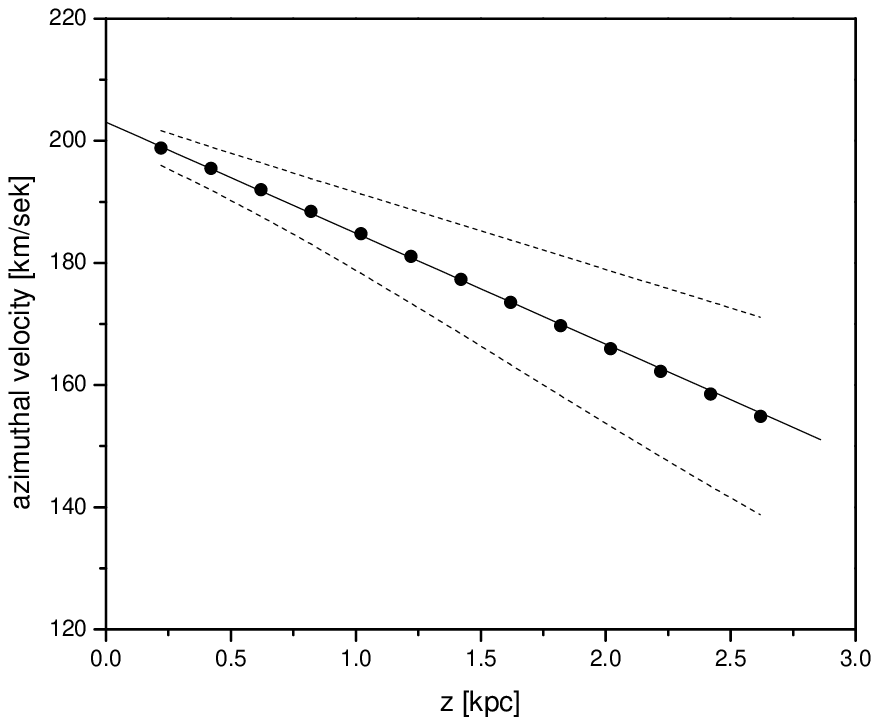}
        \includegraphics[width=0.5\textwidth]{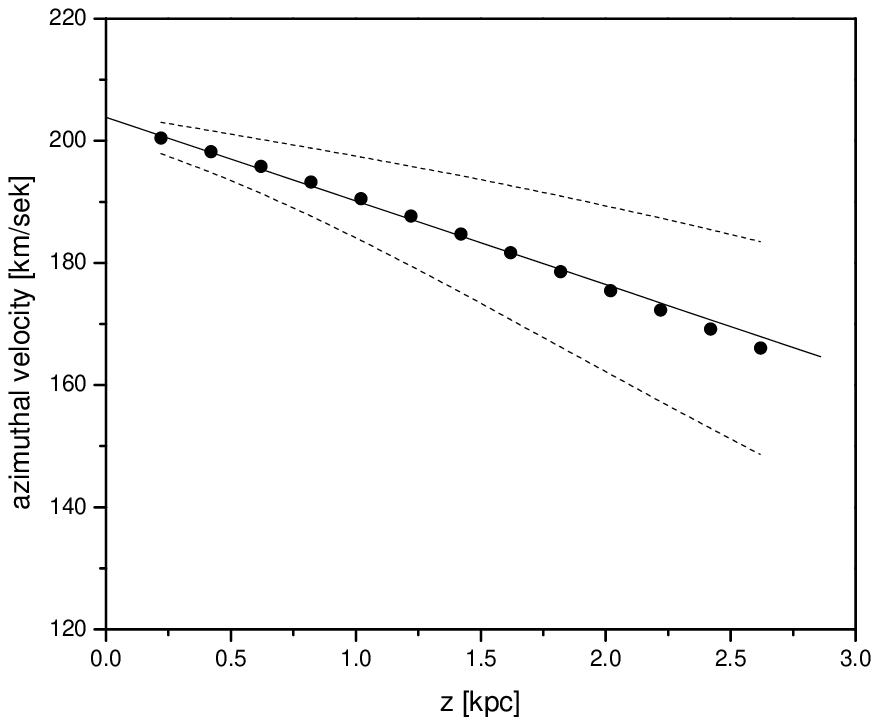}
       \caption{\label{fig:3model_gradients}
         The behaviour of the azimuthal velocity on quasi-circular orbits as a function of the distance off the Galactic mid-plane, obtained in a three-component Galaxy model considered in the text. The top figure shows the model with the smaller dark halo, whereas the bottom figure shows the model with the larger dark halo. Each point represents an average over the interval $r\in\br{3,8}\kpc$. The dashed lines represent the corresponding one-standard deviation band.}
  \end{figure}

\begin{figure}
    \centering
     \includegraphics[width=0.5\textwidth]{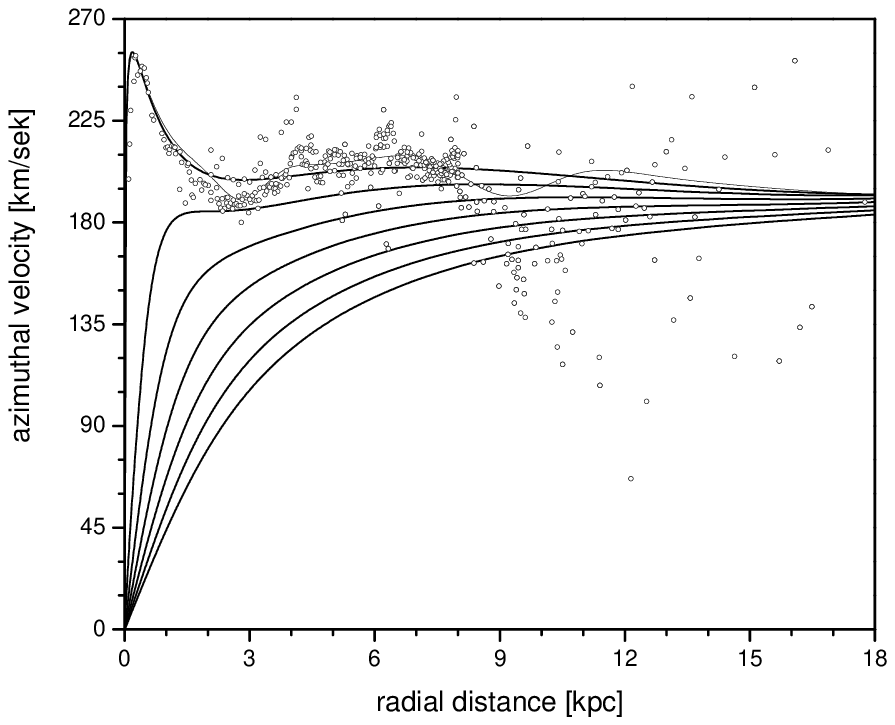}
        \includegraphics[width=0.5\textwidth]{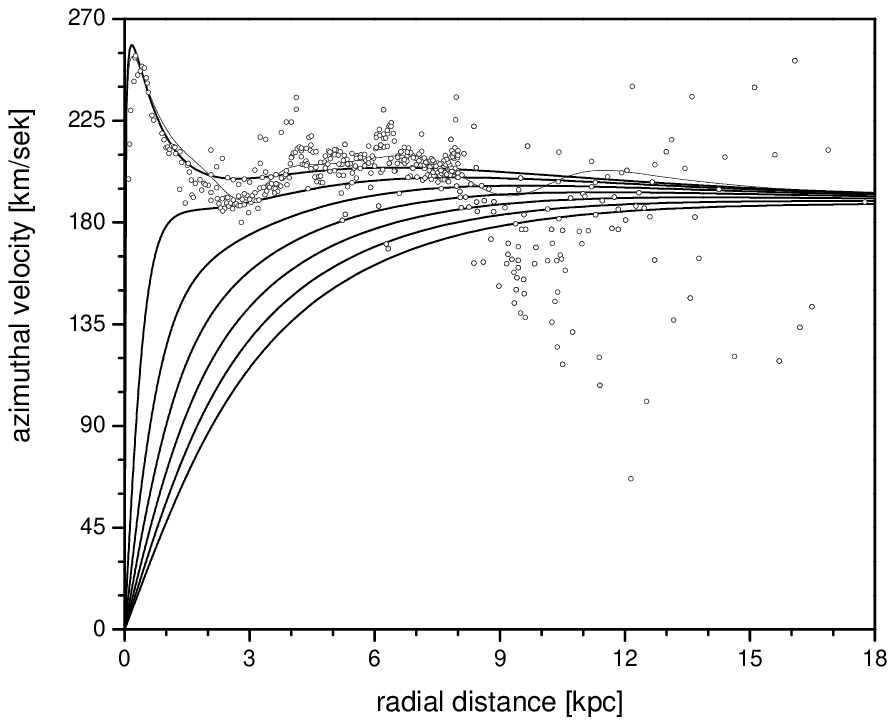}
           \caption{\label{fig:predkosci_duze_male_halo}
        Azimuthal velocity for quasi-circular orbits in a customary three-component model with the smaller dark halo (top) and with the larger dark halo (bottom) as a function of the distance from the Galactic mid-plane, shown in steps of $\Delta z=0.6\kpc$. The empty circles represent the measurements of Galaxy rotation.}
  \end{figure}

  To estimate the global gradient, we proceed in the same way as in Section \ref{sec:diskgrad2}, for the same averaging region. The averaging I-method gives $-18.2\pm3.0\gu$ and  $-13.7\pm 3.0\gu$, respectively, for models with smaller and larger dark haloes (see Fig. \ref{fig:3model_gradients}),  while the averaging II-method gives $-18.5\pm5.9\gu$  for the smaller and larger haloes, respectively. Fig. \ref{fig:predkosci_duze_male_halo} illustrates the dependence of the vertical gradient on the radial distance in the quasi-circular orbit approximation by showing rotation speeds at different heights off the Galactic mid-plane.

\subsection{Vertical gradients in the maximal halo model of the Galaxy}

 Finally, we examine how a maximum halo model deals with explaining the vertical gradients of rotation in the Galaxy. The mass function is assumed to be Keplerian,
 $M(r,z)=G^{-1}\sqrt{r^2+z^2}v_c^2\br{\sqrt{r^2+z^2}}$,
  where $v_c$ is the observed circular speed.

\begin{figure}
    \centering
       \includegraphics[width=0.5\textwidth]{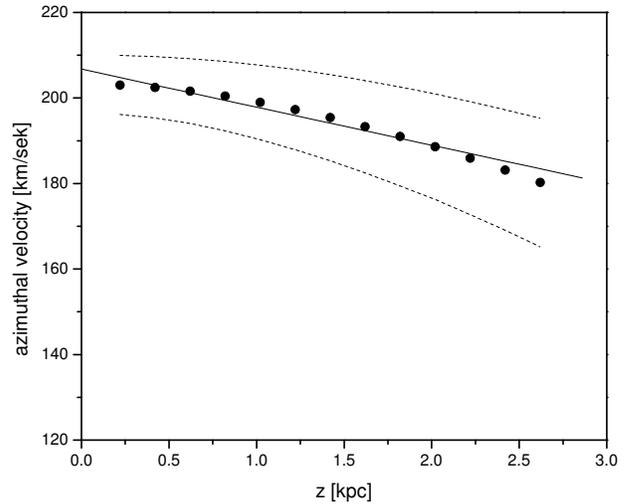}
               \caption{\label{fig:gradient_maximal_halo}
         Variation of the azimuthal velocity on quasi-circular orbits in function of the distance from the galactic midplane in the maximal halo model. Each point represents an average over the interval $r\in\br{3,8}\kpc$, the dash lines represent the corresponding 1-standard deviation band.}
  \end{figure}

\begin{figure}
    \centering
       \includegraphics[width=0.5\textwidth]{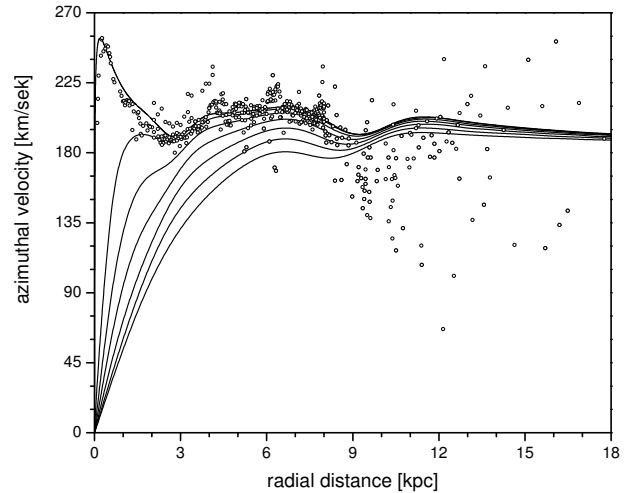}
       \caption{\label{fig:predkosci_maximal_halo}
        Azimuthal velocity  for quasi-circular orbits in function of the distance from the galaxy midplane in the maximal halo model shown in steps of $\Delta z=0.6\kpc$.}
  \end{figure}

 By analogy with the disc model symmetry, we assume the quasi-circular orbit approximation in the vicinity of the galactic mid-plane. The azimuthal component of velocity is then derived from the balance condition of the radial component of gravitational and inertial force in cylindrical coordinates,
 $r^{-1}v_{\varphi}^2=-g_r=\frac{r}{R}\frac{G M(R)}{R^2}=\frac{r}{R}\frac{R v_c^2(R)}{R^2}$,
 that is,
\[
{v}_{\varphi}(r,z)={\frac{r}{R}}v_c\br{R},\qquad R=\sqrt{r^2+z^2},\]
The vertical gradient reads
\[\nabla_z{v}_{\varphi}=\frac{r\,z}{R^2}\br{v_c'(R)-
\frac{v_c(R)}{R}},\qquad R=\sqrt{r^2+z^2}.\]
The gradient at different radii is shown in Fig. \ref{fig:gradient_analytical_disk_model}.
Again, we determine the global gradients as previously in Section \ref{sec:customary}.
The I-method
gives $-8.9\pm3.8\gu$,
see figure \ref{fig:gradient_maximal_halo}, and the II-method gives
$-9.7\pm 4.0\gu$.
Figure \ref{fig:predkosci_maximal_halo} illustrates how the vertical gradient changes with the radial variable.
The total Galaxy mass in the maximal halo model is  $1.7 \times 10^{11}\msun$ which is $1.5$ times greater than the total Galaxy mass predicted by the global disc model.
\section{Galaxy NGC 891}
In studies of the vertical gradients of rotation, the case of the Milky Way is exceptional, as the measurements have been carried out very close to the galactic mid-plane. We have already given some arguments to suggest that our gradient estimates above the galactic mid-plane could be extrapolated in the direction toward small $z$. However, to have additional support for this statement, we decided to test our approach for the edge-on galaxy NGC 891, with one of the deepest ever performed H i observations extending out to $22\kpc$ from the galaxy disc. Rotational velocity in the halo was observed to decrease in the direction normal to the galactic mid-plane, with a vertical fall-off rate of about  $-15\gu$\citep{bib:Oosterloo}. In agreement with the results are observations of diffuse ionized gas performed by \citet{bib:Benjamin}, giving a vertical gradient in azimuthal velocity in the north-east quadrant of the ionized gas halo of about  $17.5\pm5.9\gu$ ($z\in(1.2,4.8)\,\kpc$, $r\in(4.02,7.03)\,\kpc$), but with no vertical gradient detected in the south-east quadrant.

\begin{figure}
    \centering
       \includegraphics[width=0.5\textwidth]{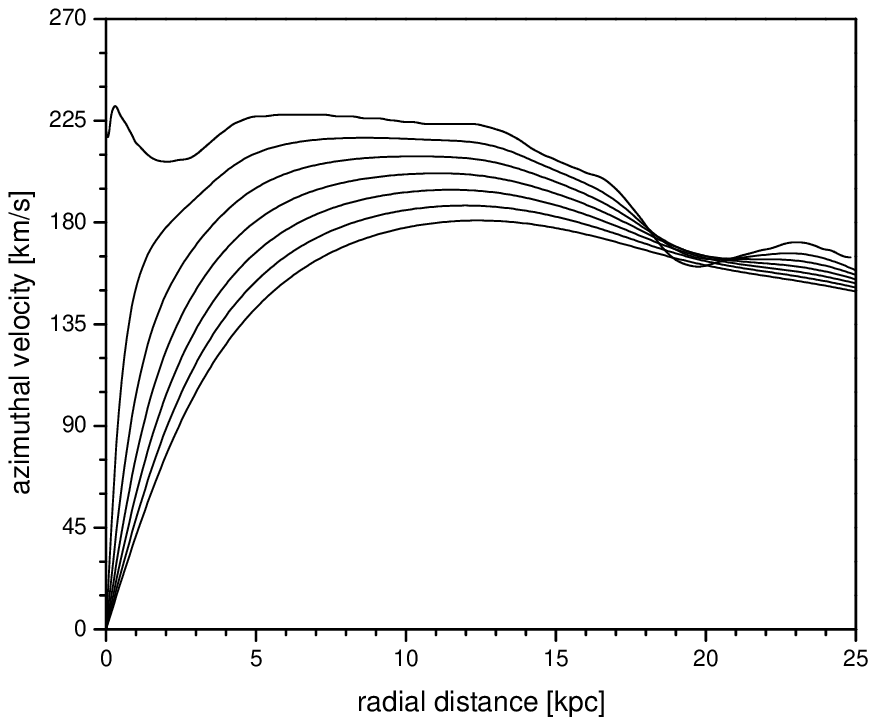}
        \includegraphics[width=0.5\textwidth]{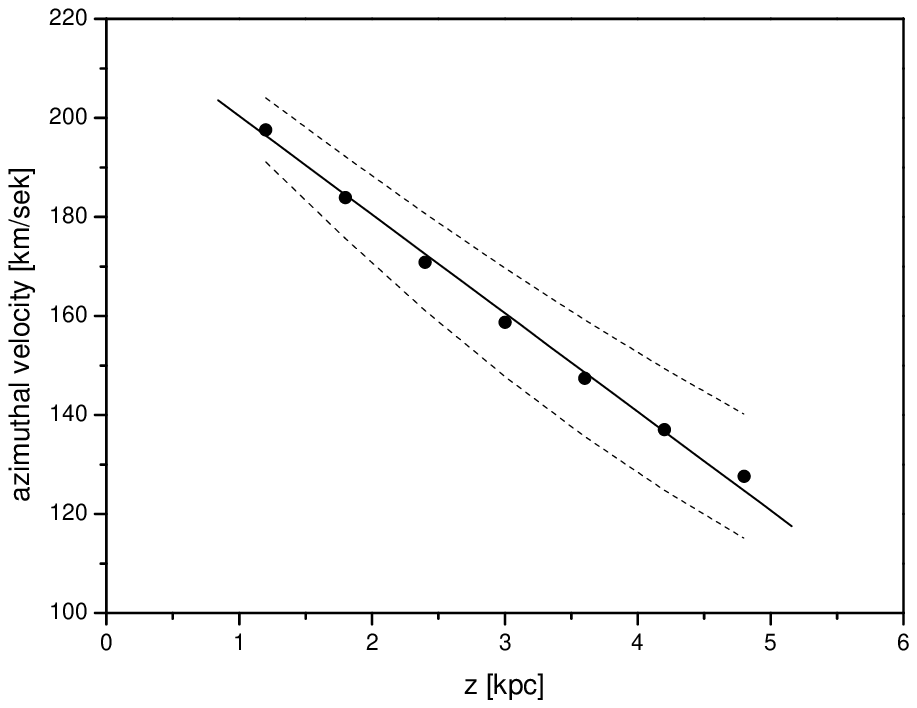}
       \caption{\label{fig:ngc891}
        Azimuthal velocity for quasi-circular orbits in function of the distance from the galactic midplane in the global disc model of galaxy NGC 891, shown in steps of $\Delta z=0.6\kpc$ (top figure) and variation of the azimuthal velocity component with the distance from the galaxy midplane (bottom figure). Each point represents an average over the interval $r\in(4,7)\kpc$, the dash lines represent the corresponding 1-standard deviation band. Rotation curve of this galaxy comes from \texttt{http://www.ioa.s.u-tokyo.ac.jp/\~{}sofue})
         }
  \end{figure}

In order to determine the global vertical gradient, we apply analytical estimates in the framework of the global thin disc model of the galaxy in the quasi-circular orbit approximation, as in Section \ref{sec:diskgrad}. This time, however, we do not study trajectories of test bodies, as we have already seen for the Galaxy that analytical estimates are sufficient, giving similar results. We make calculations in the region $z\in(1.2,4.8)\,\kpc$, $r\in(4.02,7.03)\,\kpc$.
The vertical gradients obtained for NGC 891 are  $-19.9\pm3.0\gu$ and $-19.5\pm1.7\gu$ using the I-method and II-method, respectively. These predictions overlap very well within errors with the observed vertical gradient values.

This result gives additional strong support that the thin disc idealization works well as a tool for studying vertical gradients of azimuthal velocity in flattened galaxies.

\section{Summary and conclusions}\label{sec:summary}
Throughout this paper, we have given several arguments to suggest that the global thin disc model of flattened galaxies is naturally suited for describing the large vertical gradients of rotation speed observed in the neighbourhood of the galactic mid-plane. The gradients were estimated in the quasi-circular orbit approximation, which was established in Section \ref{sec:diskgrad} to give trustworthy results.

The gradient values predicted for the Galaxy in this approximation agree very well with measurements when the disc comprises gross dynamical mass. Independent estimates of the gradient in this model (both analytical and from analysing the motion of test bodies) give consistent results. The vertical fall-off in the rotational velocity in this model is not very dependent on the height, at least out to $3\,\kpc$  above the mid-plane. This result is consistent with observations of the rotation speed in other galaxies. In other words, the distance from the mid-plane is not crucial for the gradient determination. Thus, more realistic mass models of the Galaxy, such as a finite width disc comprising the whole dynamical mass, should give similar results. We have also found that Galaxy models with significant spheroidal component are not consistent with the gradient measurements, and the discrepancy grows with the mass of the component.

With the aid of our model, we have also studied the vertical gradient in NGC 891. The gradients in this galaxy were measured at relatively large heights from the galactic disc compared to the measurements in our Galaxy. Nevertheless, we again obtained predictions in accordance with observations.

Based on all these results, we can hypothesize, contrary to what is implied by dark halo models, that gross mass distribution in our Galaxy is more flattened, disc-like, rather than spheroidal.

\end{document}